\documentclass{aa}
\usepackage{graphicx}
\usepackage{txfonts}
\usepackage{natbib}
                                                     
\def\arcmin{\hbox{$^\prime$}}
\def\arcsec{\hbox{$^{\prime\prime}$}}

\def\esc{ergs.s$^{-1}$.cm$^{-2}$}
\def\esca{ergs.s$^{-1}$.cm$^{-2}$.$\AA^{-1}$}

\begin{document}
\title{Discovery of a z = 6.17 galaxy from CFHT and VLT observations
\thanks{Based on observations collected at the Canada-France-Hawaii Telescope 
operated by the National Research Council of Canada, the Centre National de 
la Recherche Scientifique de France and the University of Hawaii 
(Programme 01BF1) and on observations collected at the European Southern 
Observatory, Paranal, Chile (ESO Programme 70.A-0513)}}

   \subtitle{}

   \author{J.-G. Cuby\inst{1}
          \and
          O. Le F\`evre\inst{2}
	  \and
	  H. McCracken\inst{3}
          \and
	  J.-C. Cuillandre\inst{4}
	  \and
	  E. Magnier\inst{4}
	  \and
	  B. Meneux\inst{2}
}

   \offprints{J.-G. Cuby}

   \institute{European Southern Observatory, 
Paranal, Casilla 19001, Santiago 19, Chile \email{jcuby@eso.org}
\and Laboratoire d'Astrophysique de Marseille (France)
\and Osservatorio Astronomico di Bologna (Italy)
\and Canada-France-Hawaii Telescope Corporation (Hawaii, USA)
}

   \date{Received <date> / Accepted <date>}

   \abstract{We report the discovery of a galaxy at a redshift z = 6.17
identified from deep narrow band imaging and spectroscopic follow-up
in one of the CFHT-VIRMOS deep survey fields at 0226-04. In addition
to the existing deep BVRI images of this field, we obtained a very deep 
narrow band image at 920 nm with the aim of detecting Ly$\alpha$ emission 
at redshift $\sim$ 6.5.
Spectroscopic follow-up of some of the candidates selected on the basis
of their excess flux in the NB920 filter was performed at the 
VLT-UT4 with the FORS2 instrument. For one object a strong and asymmetric 
emission line associated with a strong break in continuum
emission is identified as Ly$\alpha$ at $z$ = 6.17. 
This galaxy was selected from its continuum emission in the
920 nm filter rather than for its Ly$\alpha$ emission, in effect
performing a Lyman Break detection at z = 6.17.
We estimate a star formation 
rate of several tens of M$_\odot$.yr$^{-1}$ for this object,
with a velocity dispersion $\sim$ 400 km.s$^{-1}$.
The spectroscopic follow-up of other high $z$ galaxy candidates
is on-going.

   }

   \maketitle
%

\section{Introduction}
Recent observations of galaxies at redshifts $>$ 5, and more 
recently $>$ 6 have been reported. These observations follow a
well defined observing strategy based on Ly$\alpha$ searches at
increasing wavelengths in low OH emission sky windows 
\citep[see e.g.][]{hu02,cuby02,rhoads03,kodaira03,lilly03} and / or on Lyman
Break detection in the RIz bands \citep[e.g.][]{lehnert03}. 
The Sloan Digital Sky Survey has in the meantime unveiled a large number of 
z $>5$ QSO's \citep{fan99,anderson01} from Lyman break detection.

These recent observations have generated extensive
debates in the literature regarding a possible redshift of the 
re-ionization of the Universe at z $\approx$ 6 
(see e.g. \citet{becker01,hu02,lehnert03} and references therein). 
In February 2003 the publication
of the WMAP results \citep{bennet03} has shed new light on the early epochs of the Universe,
in particular in dating the epoch of the re-ionization at $11 < z < 30$.
This suggests that the current observational limit at z $\approx$ 6.5 corresponds
to an epoch significantly beyond the re-ionization period and that previous 
failures and recent success in detecting z $>$ 6 objects
are due to the evolution of the search techniques more than to
the ionization state of the IGM at this redshift. 
However various models of the re-ionization phases
of the Universe by massive stars, miniquasars or neutrinos 
allow to reconcile the z $\sim$ 6 Gunn-Peterson trough observations with the WMAP data
\citep{Haiman03,Hansen03,cen03}. The detectability of the Ly$\alpha$ emission line for
a galaxy embedded in a neutral IGM is discussed in \citet{Haiman02}.

We report in this letter the discovery of a z = 6.17 galaxy
discovered from Narrow Band imaging performed at
CFHT and follow-up spectroscopy at the VLT. 
In section~\ref{sec:observations} we present the observations and
in section~\ref{sec:results} the results which are briefly 
discussed in section~\ref{sec:discussion}.

We assume in the following a cosmology based on the most recent
values derived from WMAP data, i.e. H$_{\rm o}$ =
71 km.s$^{-1}$.Mpc$^{-1}$, $\Omega_{matter}$ = 0.27 and a
flat Universe.

\section{Observations}
\label{sec:observations}
\subsection{Imaging and Object Selection}
We have obtained deep Narrow Band imaging at 920 nm with the CFH12K 
CCD mosaic \citep{cuillandre00}
during 4 nights in November 2001. The selected field at 0226-04 
is one of the VIRMOS-VLT Deep Survey (VVDS) 
fields \citep{lefevre01} for which deep photometric catalogs from 
broad band BVRI data obtained with the same instrument were 
already available. The existing BVRI data 
reach limiting magnitudes of I$_{\rm AB}$=25.3 for a $3\sigma$ detection
in a 3 arcsecond circular aperture \citep{mccracken03}.
The field of view is 28\arcmin $\times$ 42\arcmin\ with 0.2\arcsec\ sampling.
Conditions were photometric. In total 20 hrs of integration were accumulated 
at a median airmass of 1.22 with individual exposure times of 900 s. 

The NB920 filter was designed with a central wavelength
920 nm and a 10 nm bandwidth to match a region of 
low sky emission between OH lines. The measured background level was
$\sim$ 1 e$^-$.s$^{-1}$.pixel$^{-1}$  providing background limited 
observations. The resulting image quality on the combined image is $\sim$ 0.8\arcsec\
with some slight image elongation ($\sim$ 10\%) due to oscillations of the declination axis
of the telescope.

The 80 dithered images were centered and stacked after standard pre-processing: 
bad pixel masking, dark correction and (twilight) flat-fielding and fringe subtraction.
The stacked image was then distortion corrected,  astrometrically calibrated
and matched to the BRVI images of the field. Some residual fringing persisted,
which was further removed chip by chip by a 50 order spline 3 fitting.

Object detection on the NB920 image was then performed chip by chip 
with SExtractor \citep{bertin96}. Careful visual inspection allowed 
to further remove false detections, in particular in the regions
of poor cosmetic quality and / or of overlap between the chips of the
mosaic. For all remaining objects the photometry
of the $\chi^2$ BVRI VIRMOS detection image  \citep{mccracken03}
was measured using DAOPHOT under IRAF. 
The final selection was performed
by keeping all objects for which the BVRI flux was below an arbitrary
low level. Thumbnails of the fields around each candidate were extracted 
and visual inspection further allowed to remove remaining false and / or
ambiguous detections.

\subsection{Spectroscopy}
Spectra of some candidates were obtained at the VLT with 
FORS2 using the 19 movable slitlet capability of the
instrument, with slit widths of 2\arcsec\ and a pixel
scale of 0.25\arcsec. The seeing conditions were in average of the
order of 0.8--1.0\arcsec. In total, 3 fields and approximately 30 candidates were 
observed in a combination of service and visitor mode observations
from November 2002 to January 2003.
The combination of a MIT/LL CCD mosaic with a holographic
grism provides extremely good performance
up to 1 $\mu$m with little fringing. The setup used provides
a reciprocal
dispersion of 1.6 $\AA$.pixel$^{-1}$ giving a spectral resolution 
in the range 700--1400 depending on the size of the objects through 
the 2\arcsec\ slits.

Observations on each field consisted of 1.9 hr of integration time
split in 6 exposures with dithering along the slits.
After flat fielding, running skies were generated for
each frame from the 5 other frames. The sky subtracted frames were then 
distortion corrected, registered and stacked, and sky line residuals
were removed by column fitting.

One target was clearly identified on the resultant 2D combined images 
as a high redshift galaxy by the presence of a strong, asymmetric emission 
line and a strong continuum break across the line. 
The total spectral coverage for this object is 762--1091 nm.
The emission line lies at shorter wavelength than in the expected 
920 nm window: the relatively strong continuum caused instead this object to be detected
in the NB920 image and be selected for its continuum Lyman break.

Results for this particular object are presented
in this letter, while the analysis of the 
imaging and spectroscopic data for the full
sample will be presented in a subsequent paper.

\section{Results}
\label{sec:results}
\subsection{Imaging}

\begin{figure}
\resizebox{\hsize}{!}{\includegraphics{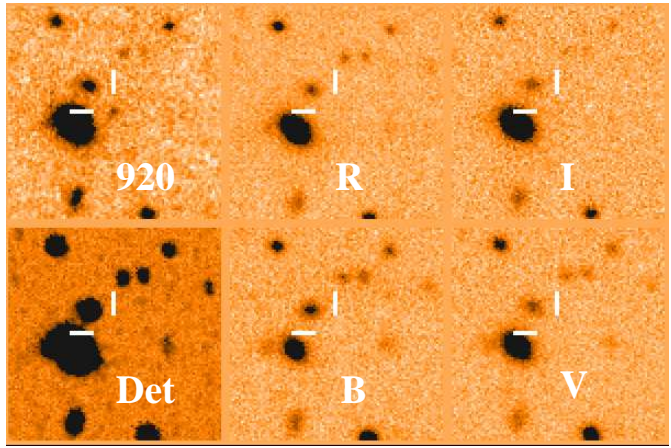}}
\caption{Thumbnail images of the field of view in BVRI, 920 nm images 
around the target. 'Det' indicates the $\chi^2$ co-added detection image 
constructed from the BVRI images of the field \citep{mccracken03}. 
The thumbnails are 20\arcsec\ on a side. East is right, North is up.}
\label{fig:thumbnails}
\end{figure}

Figure~\ref{fig:thumbnails} shows BVRI and NB920 thumbnail images around
the object. The object is unresolved at the spatial resolution of the 
image ($\sim$ 0.8\arcsec).

Flux calibration was performed on a number of standard fields, SA111, SA113
and MarkA. In total 9 stars were used in the 3 fields to derive the
photometric zero point at 920 nm. Furthermore, 2 of these 3 fields were
dithered across the mosaic, allowing to characterize the sky 
concentration and scattered light patterns and to construct 
a photometric superflat correction map.
The 920 nm magnitudes were derived
by an ad'hoc extrapolation of the BVRI magnitudes of the 9 standard stars
leading to I band magnitude corrections of $\pm$ 0.1 to 0.5. 
In spite of the uncertainties associated to this process,
the resulting statistical errors turned out to be
$<$ 0.1 magnitude peak to peak. The measured magnitude for the object
is NB920 = 23.77, corresponding to a flux density of $\sim$ 2.6 10$^{-19}$ \esca.

\subsection{Spectroscopy}
\begin{figure*}
\centering
\includegraphics[angle=-90,width=17cm]{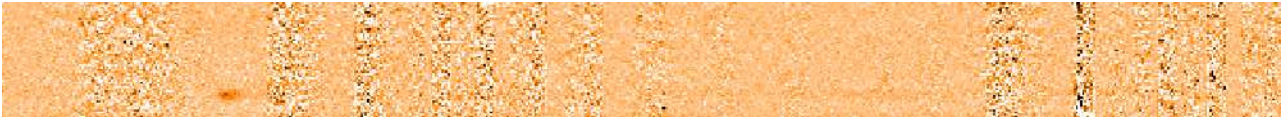}
\caption{2D spectrum. The emission line is clearly visible, with its asymmetric
profile, and the continuum is visible throughout the 920 nm  OH free window.
Wavelength increases linearly to the right from 857 to 955 nm.}
\label{fig:2Dspectrum}
\end{figure*}

Figure~\ref{fig:2Dspectrum} shows the 2D spectrum of the galaxy as
observed with FORS2. 
The total absence of continuum detection 
blueward of the line is consistent with the non detection of the 
object in the R
image. The non detection in the I band image is further discussed below.

\begin{figure}
\resizebox{\hsize}{!}{\includegraphics{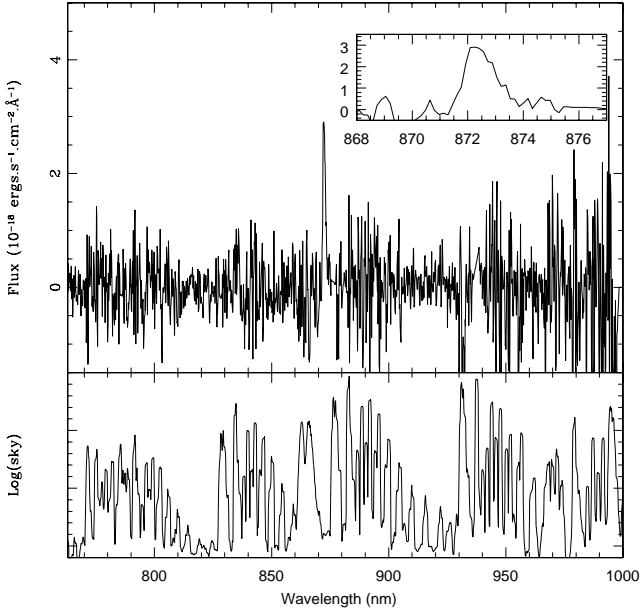}}
\caption{Extracted spectrum in the range [760-1000] nm, 
with a close-up view of the emission line in the range [868-877] nm.
The sky spectrum is shown in  arbitrary logarithmic units.}
\label{fig:1Dspectrum}
\end{figure}

Flux calibration was done from observation of a standard star (LTT3218)
observed through a 5\arcsec\ aperture located at the center of the field of 
view of the instrument and covering the spectral range 731-1059 nm. The
spectral range common to the object and the standard star is therefore
[762--1059] nm.
The resulting flux calibrated spectrum is shown on figure~\ref{fig:1Dspectrum}.


The line peaks at 872.3 nm. We measure a line intensity of  3.9 10$^{-17}$ \esc\ with 
a cumulated error (slit losses, spectrophotometric calibration,
measurement errors and signal to noise) estimated to be less than 
20\% (3$\sigma$). 

The continuum, although at a S/N $\sim$ 1 per pixel could be 
estimated with good accuracy by fitting the spectroscopic 
2D frame (figure~\ref{fig:2Dspectrum}) along the dispersion direction
in selected low noise windows (before and after the line).

We measure a mean flux density of $0. \pm 3~10^{-20}$ \esca
(3 $\sigma$)
in the range 809--828 nm and $\sim2.4 \pm 0.6~10^{-19}$
in the range 914--928 nm (corresponding to the NB filter), in excellent
agreement with the CFH12k NB920 photometry.
The amplitude of the continuum break across the emission line is
therefore $>$ 8 or $>$ 2.3 magnitudes.

We have examined several possibilities 
for the emission line: H$\alpha$ at z=0.329, [OII] at z=1.340 or z=6.17 
for Ly$\alpha$. The identification with Ly$\alpha$ is the most likely on the 
basis of the following arguments.

First, the strong continuum break across the line is only compatible
with Ly$\alpha$. 
We have examined a complete set of simulated spectra produced
with GISSEL \citep{charlot93}, spanning a range of burst age
from 0.01 to 12 Gyr, and metallicities from 0.004
to solar. The amplitude of the continuum break never
exceeds 1.8 magnitude for [OII] and only for models with 
old stellar populations unlikely to exhibit [OII] in
emission, it is 0.2--0.3 magnitude around
H$\alpha$ while it can easily exceed 2-3 magnitudes around
Ly$\alpha$.

Second, the equivalent width at rest also favors identification with
Ly$\alpha$. From the averaged continuum under the  line 
we derive rest frame equivalent
widths of $\sim$ 50 $\AA$ for Ly$\alpha$, 150 $\AA$ for [OII] and
270 $\AA$ for H$\alpha$. A value of 150 $\AA$ is at the extreme
end of the distribution for [OII] at redshifts $\sim1$ \citep{hammer97},
a value of 270 $\AA$ has never been observed for H$\alpha$ at 
low redshifts \citep{tresse96}, while the value for Ly$\alpha$ is
quite common at high redshifts \citep{steidel99,hu02,rhoads03}.

Finally, the line is clearly asymmetric (see figure 
\ref{fig:1Dspectrum}), as typical of Ly$\alpha$ at high redshifts
with a sharp fall-off in the blue due to interstellar and / or IGM absorption.

We conclude that the most likely identification 
of the observed emission line is Ly$\alpha$ at $z~\sim~6.17$,
by slightly underestimating the redshift derived 
from the line peak to account for a likely typical distortion 
of the Ly$\alpha$ line on its blue side.

The observed line width is 13 $\AA$. Since the object is not resolved
in imaging (in the UV continuum above Ly$\alpha$), we assume
that the instrumental profile is given by the seeing disk,
i.e. $\sim$ 0.8\arcsec\
as measured across the trace of bright objects in the same frame.
This leads to a velocity dispersion $\sim$ 400 km.s$^{-1}$.

From the flux measurements above, 
we can verify a posteriori the integrated
I band magnitude for this object. Assuming that the continuum level
is constant above the Ly$\alpha$ line and nil below, and taking
into account the decrease in CCD quantum efficiency across the I band
we derive that our object should have a magnitude I$_{\rm AB}~\sim~$25.9
through the I filter of CFH12k, magnitude for which
the completeness limit of our I band image is $\sim$ 50\%
\citep{mccracken03} and this confirms a posteriori that this object 
could indeed be below the detection threshold in the I band 
(see figure~\ref{fig:thumbnails}).

The measured properties of our object are summarized in table~\ref{tab:properties}.

\begin{table*}
\begin{center}
\centering
\caption{\label{tab:properties}Properties of our 6.17 galaxy. $^{[1]}$: rest frame, $^{[2]}$: observed frame}
\begin{tabular}{llccccccc} \hline
Alpha & Delta & NB920 & Ly$\alpha$$^{[2]}$ & \multicolumn{2}{c}{FWHM} & EW$^{[1]}$ & $F$(Ly$\alpha$)$^{[2]}$ & $F$(127.5.--129.4 nm)$^{[2]}$ \\ 
(2000)& (2000)& mag.  &  nm        & nm$^{[2]}$    & km.s$^{-1}$ & $\AA$ & 10$^{-17}$ \esc       &10$^{-19}$ \esca  \\ \hline

02:28:02.97 & -04:16:18.3 & 23.77 & 872.3 & 1.3 &400  & 50 & 3.9 & 2.5 \\ \hline
\end{tabular}
\end{center}
\end{table*}

\section{Discussion}
\label{sec:discussion}
Observation of objects at the highest
redshifts is critical for our understanding of the formation
and evolution of the galaxies and of the early phases of the 
Universe. Although we examine only one such object, it is of interest to
assess its properties.

Classical estimators of the star formation rates \citep{kennicutt98}
from the Ly$\alpha$
luminosity (assuming case B recombination) and UV continuum flux
(although used here at shorter wavelengths than the recommended
1500--2800
$\AA$ range) give for our object and for our assumed cosmology
SFR(Ly$\alpha$) = 15 M$_\odot$.yr$^{-1}$ and SFR(UV) = 60
M$_\odot$.yr$^{-1}$.
The higher rate SFR(UV), albeit similar to SFR(Ly$\alpha$) within
the uncertainties of the estimators, is not unusual and could reflect the
effects of dust absorption on the Ly$\alpha$ photons in the interstellar
and intergalactic medium. This SFR is similar to the values
derived from other high-z galaxy observations 
\citet{kodaira03,hu02,rhoads03}.

Our measurements indicate a strong velocity field of several hundreds of 
km.s$^{-1}$, which might be related to the inflow of 
gas onto a recently formed, or being assembled, galaxy. As for the
brightness of the line, it is however important to note that this value
is a lower limit of the actual line width.

All existing observations of high redshift galaxies indicate moderate to 
vigorous star formation rates.
The WMAP dating of the re-ionization of the Universe \citep{kogut03}
indicates that the first stars were born very early in the Universe, before
the first 400 Myr. Objects $\sim$ 0.9 Gyr old (z $\sim$ 6.5) may therefore have 
already experienced several cycles of star formation and metal enrichment,
as observed in z $>$ 6 SDSS quasars \citep{pentericci02} and 1.5 Gyr later 
at redshift 2.7 in a Lyman break galaxy \citep{pettini02}. Conversely,
the high EW of Ly$\alpha$ in the high-z galaxies observed so far rather
point to low metallicity objects. The strong Ly$\alpha$ luminosity of $\sim$ 1.7 
10$^{43}$ ergs.s$^{-1}$ determined for our object may imply a young object
\citep{lehnert03}.

Short of higher S/N spectroscopy at higher wavelengths,
it is difficult to infer the metallicity of the few high z galaxies known.
Clearly more zJHK observations are required to better determine the nature and
chemical composition of these distant galaxies, and to compare them with
quasars at the same redshifts or LBG's at lower redshifts.

We defer to a later paper the analysis of the space density of high redshift
galaxies based on the CFHT and VLT-FORS2 observations of our complete
sample of NB920 candidates.

\section{Conclusion}
We report in this letter the discovery of a field galaxy at redshift
6.17 as part of a dedicated narrow band imaging search for z $\sim$ 6.5 
galaxies. We note that our object was discovered through its UV continuum
emission and not through its Ly$\alpha$ emission, indicating that 
deep imaging on 4m class telescopes can indeed turn
out z$\sim$ 6 objects from continuum detection.

Similar dedicated high-z galaxy searches are now consistently reporting
discoveries of z $>$ 6 objects, i.e. less 1 Gyr old objects, undergoing
moderate to high star formation.

Larger samples are required to understand how galaxies evolve through
this important epoch in the life of the Universe.

\begin{acknowledgements}
We thank S. Arnouts for providing the GISSEL simulations mentioned in text, 
the anonymous referee for useful comments,
the CFHT and ESO staff at Paranal and Garching for their always 
efficient and dedicated support and the time allocating committees for 
granting us observing time.
\end{acknowledgements}


%


\bibliographystyle{aa}
\bibliography{references2}

\begin{thebibliography}{26}
\expandafter\ifx\csname natexlab\endcsname\relax\def\natexlab#1{#1}\fi

\bibitem[{Anderson {et~al.}(2001)Anderson, Fan, Richards, Schneider, Strauss,
  Berk, Gunn, Knapp, Schlegel, Voges, \& Yanny}]{anderson01}
Anderson, S., Fan, X., Richards, G., {et~al.} 2001, AJ, 122, 503

\bibitem[{Becker {et~al.}(2001)Becker, Fan, White, Strauss, Narayanan, Lupton,
  Gunn, Annis, Bahcall, Brinkmann, Connolly, Csabai, Czarapata, Doi, Heckman,
  Hennessy, Ivezic, Knapp, Lamb, McKay, Munn, Nash, Nichol, Pier, Richards,
  Schneider, Stoughton, Szalay, Thakar, \& York}]{becker01}
Becker, R., Fan, X., White, R., {et~al.} 2001, AJ, 122(6), 2850

\bibitem[{Bennett {et~al.}(2003)Bennett, Halpern, Hinshaw, Jarosik, Kogut,
  Limon, Meyer, Page, Spergel, Tucker, Wollack, Wright, Barnes, Greason, Hill,
  Komatsu, Nolta, Odegard, Peirs, Verde, \& Weiland}]{bennet03}
Bennett, C., Halpern, M., Hinshaw, G., {et~al.} 2003, submitted to ApJ,
  astro-ph/0302207

\bibitem[{Bertin \& Arnouts(1996)}]{bertin96}
Bertin, E. \& Arnouts, S. 1996, A\&AS, 117, 393

\bibitem[{Bruzual \& Charlot(1993)}]{charlot93}
Bruzual, A. \& Charlot, S. 1993, ApJ, 405, 538

\bibitem[{Cen(2003)}]{cen03}
Cen, R. 2003, submitted to ApJ, astro-ph/0210473

\bibitem[{Cuby {et~al.}(2002)Cuby, \mbox{Le~F\`evre}, McCracken, Cuillandre,
  Magnier, Gilmozzi, Moorwood, \& van~der Werf}]{cuby02}
Cuby, J.-G., \mbox{Le~F\`evre}, O., McCracken, H., {et~al.} 2002, in Proc.
  SPIE, Vol. 4834,

\bibitem[{Cuillandre {et~al.}(2000)Cuillandre, Luppino, Starr, \&
  Isani}]{cuillandre00}
Cuillandre, J.-C., Luppino, G., Starr, B., \& Isani, S. 2000, in Proc. SPIE,
  Vol. 4008, , 1010

\bibitem[{Fan {et~al.}(1999)Fan, Strauss, Schneider, Gunn, Lupton, Yanny,
  Anderson, Anderson, J.~Annis, Bahcall, Bakken, Bastian, Berman, Boroski,
  Briege, \& Briggs}]{fan99}
Fan, X., Strauss, M., Schneider, D., {et~al.} 1999, AJ, 118, 1

\bibitem[{Haiman(2002)}]{Haiman02}
Haiman, Z. 2002, ApJ, 576, L1

\bibitem[{Haiman \& Holder(2003)}]{Haiman03}
Haiman, Z. \& Holder, G. 2003, submitted to ApJ, astro-ph/0302403

\bibitem[{Hammer {et~al.}(1997)Hammer, Flores, Lilly, Crampton, F\`evre, Rola,
  Mallen-Ornelas, Schade, \& Tresse}]{hammer97}
Hammer, F., Flores, H., Lilly, S., {et~al.} 1997, ApJ, 481, 49

\bibitem[{Hansen \& Haiman(2003)}]{Hansen03}
Hansen, S. \& Haiman, Z. 2003, submitted to ApJ, astro-ph/0305126

\bibitem[{Hu {et~al.}(2002)Hu, Cowie, McMahon, Capak, Iwamuro, Kneib, Maihara,
  \& Motohara}]{hu02}
Hu, E., Cowie, L., McMahon, R., {et~al.} 2002, ApJ, 568(2), L75

\bibitem[{Kennicutt(1998)}]{kennicutt98}
Kennicutt, R. 1998, ARA\&A, 36, 189

\bibitem[{Kodaira {et~al.}(2003)Kodaira, Taniguchi, Kashikawa, Kaifu, Ando,
  Karoji, Ajiki, Akiyama, Aoki, Doi, Fujita, \& Furusawa}]{kodaira03}
Kodaira, K., Taniguchi, Y., Kashikawa, N., {et~al.} 2003, PASJ, 55

\bibitem[{Kogut {et~al.}(2003)Kogut, Spergel, Barnes, Bennett, Halpern,
  Hinshaw, Jarosik, Limon, Meyer, Page, Tucker, Wollack, \& Wright}]{kogut03}
Kogut, A., Spergel, D., Barnes, C., {et~al.} 2003, submitted to ApJ,
  astro-ph/0302213

\bibitem[{Lehnert \& Bremer(2003)}]{lehnert03}
Lehnert, M. \& Bremer, M. 2003, Accepted for publication to Ap.J.,
  astro-ph/0212431

\bibitem[{Lilly {et~al.}(2003)Lilly, Tran, Brodwin, Crampton, Juneau, \&
  McCracken}]{lilly03}
Lilly, S., Tran, K.-V., Brodwin, M., {et~al.} 2003, submitted to ApJ,
  astro-ph/0304376

\bibitem[{\mbox{Le F\`evre} {et~al.}(2001)\mbox{Le F\`evre}, Vettolani,
  Maccagni, Mancini, Mazure, Mellier, Picat, Arnaboldi, Bardelli, Bertin,
  Busarello, Cappi, \& Charlot}]{lefevre01}
\mbox{Le F\`evre}, O., Vettolani, G., Maccagni, D., {et~al.} 2001, in Deep
  Fields, ed. S.~Cristiani, A.~Renzini, \& R.~Williams, Proceedings of the
  ESO/ECF/STScI Workshop, 236

\bibitem[{McCracken {et~al.}(2003)McCracken, Radovich, Bertin, Mellie,
  Dantel-Fort, F\`evre, Cuillandre, Gwyn, Foucaud, \& Zamorani}]{mccracken03}
McCracken, H., Radovich, M., Bertin, E., {et~al.} 2003, Accepted for
  publications to A\&A

\bibitem[{Pentericci {et~al.}(2002)Pentericci, Fan, Rix, Strauss, Narayanan,
  Richards, Schneider, Krolik, Heckman, Brinkmann, Lamb, \&
  Szokoly}]{pentericci02}
Pentericci, L., Fan, X., Rix, H.-W., {et~al.} 2002, AJ, 2158, 123:2151

\bibitem[{Pettini {et~al.}(2002)Pettini, Rix, Steidel, Adelberger, L., Hunt, \&
  Shapley}]{pettini02}
Pettini, M., Rix, S., Steidel, C., {et~al.} 2002, ApJ, 569, 742

\bibitem[{Rhoads {et~al.}(2003)Rhoads, Dey, Malhotra, Stern, Spinrad, Jannuzi,
  Dawson, Brown, \& Landes}]{rhoads03}
Rhoads, J., Dey, A., Malhotra, S., {et~al.} 2003, AJ, 125, 1006

\bibitem[{Steidel {et~al.}(1999)Steidel, Adelberger, Giavalisco, Dickinson, \&
  Pettini}]{steidel99}
Steidel, C., Adelberger, K., Giavalisco, M., Dickinson, M., \& Pettini, M.
  1999, ApJ, 519, 1

\bibitem[{Tresse {et~al.}(1996)Tresse, Rola, Hammer, Stasinska, F\`evre, Lilly,
  \& Crampton}]{tresse96}
Tresse, L., Rola, C., Hammer, F., {et~al.} 1996, MNRAS, 281, 847

\end{thebibliography}

\end{document}